# PMMA-grafted graphene nanoplatelets to reinforce the mechanical and thermal properties of PMMA composites

Cristina Vallés[*], Dimitrios G. Papageorgiou, Fei Lin, Zheling Li, Ben F. Spencer, Robert J. Young, Ian A. Kinloch[*]

School of Materials and National Graphene Institute, University of Manchester, Oxford Road, Manchester, M13 9PL, UK

**Abstract**

In order to realise the potential of graphene nanocomposites it is vital to control the degree of dispersion and achieve a strong graphene/polymer interface. Herein, we developed a facile 'grafting to' functionalisation approach for graphene nanoplatelets. $NH_2$-terminated graphene nanoplatelets ($NH_2$-GNPs) prepared by a diazonium coupling were used as a 'platform' to covalently graft PMMA chains to the surface of graphene through an amidation between the $–NH_2$ groups and PMMA chains (PMMA-NH-GNPs). A degree of PMMA grafting of ~3.8 wt.% (one chain per ~40 carbon atoms) was found to both improve the dispersion of the GNPs in a PMMA matrix and give strong graphene/polymer interfaces compared to as-provided GNPs. Thus, 2 wt.% of PMMA-NH-GNPs in PMMA was found to increase the elastic modulus, strength and strain at break of PMMA, whereas the incorporation of unmodified GNPs showed poor levels of reinforcement at all loadings. Furthermore, $T_g$ and $T_d$ of PMMA were increased by 15 °C and 29 °C, respectively, by adding 5 wt.% of PMMA-NH-GNPs, whereas incorporating unmodified GNPs led to smaller increases. This work offers the possibility of controlling the properties of graphene/polymer composites through chemically tuning the graphene/polymer interface, which will have broad implications in the field of nanocomposites.

[*]Corresponding authors.
cristina.valles@manchester.ac.uk; Tel: +44(0)161 306 1454
ian.kinloch@manchester.ac.uk; Tel: +44(0)161 306 3615

**1. Introduction**

Since its isolation, graphene has attracted strong interest for a range of applications, including composite materials, transistors, electronics, supercapacitors, hydrogen storage and solar cells [1-3]. Graphene-related materials (GRMs) are of particular interest to the field of polymer composites due to GRMs' exceptional mechanical, electrical and thermal properties. However, to achieve this potential, good dispersions of the flakes in the polymer matrix and strong interfaces between both components are required. Gong *et al*. [4] used Raman spectroscopy to monitor the stress transfer efficiency and breakdown of the graphene/polymer interface, finding that the graphene/polymer stress transfer depended on both $s$ (*i.e.* aspect ratio of the graphene) and the parameter $n$ (widely accepted as an effective measure of the interfacial stress transfer efficiency, *i.e.* the degree of interaction that graphene has with the polymer matrix). They probed that the interface between unfunctionalized graphene and polymer is very weak and $ns > 20$ is required for a sufficient reinforcement of the polymer. Relatively large graphene flakes will be thus needed before efficient reinforcement can take place in a graphene/polymer composite ($> 20$ µm, as we probed experimentally in PMMA and PP matrices [5, 6]). Alternatively, chemical modification of the surface or edges of the graphene flakes may significantly strengthen the interface between graphene and polymer, reducing the critical length (thus $s$) and increasing $n$. We recently showed that graphene oxide (GO) has a stronger interface with poly(methyl methacrylate) (PMMA) relative to un-functionalized graphene. This strong interface was reasoned to be due to the presence of the high number of oxygen-containing functionalities on the surface of the GO flakes [6, 7]. However, it was found that at loadings above 1 wt.% the flake-flake interactions began to dominate over flake-polymer interactions, leading to the formation of agglomerates through π-π and van der Waals interactions. These aggregates inevitably caused the deterioration of the mechanical performance of the final composite. The controlled grafting of functional groups to the graphene related materials (GRM) surfaces could prevent the agglomeration of graphene sheets through steric stabilisation, whilst still promoting a strong filler/polymer interaction, which will promote enhanced mechanical and thermal properties for the composites.

The grafting of aryl groups by reduction of the corresponding diazonium salts, previously used for carbon nanotubes (CNTs), pyrolytic graphite, glass carbon electrodes, etc. [8, 9], has been recently applied to GO-derived materials [10-12] with the objective of improving the thermal and mechanical properties of polymer composites through enhanced compatibilities between filler and polymer. However, in order to achieve stable dispersions of graphene and optimize the microstructure of the nanocomposites, polymer functionalization of the graphene surface is often necessary, particularly when non-polar polymers are used as the host matrix [13]. The incorporation of polymer grafted GO materials into host polymers of different nature has been reported to improve their compatibilities and filler/polymer interfaces, leading to enhanced mechanical and/or thermal properties of the nanocomposites. For example, the incorporation of polyetheramine-functionalized GO into an epoxy resin was reported to render significant enhancements of the mechanical properties of the epoxy (up to 63% in tensile strength and 90% in toughness for 0.5 wt.% filler loading), as well as of the thermal properties (the glass transition temperature, $T_g$, of the polymer was found to increase with the addition of this filler) [14]. Epoxy-grafted-GO was recently shown to form chemical bonds with a polycarbonate (PC) matrix, leading to increases of the Young's modulus and tensile strength by 52.3% and 5.3%, respectively, for 0.5 wt.% loaded GO-epoxy/PC composite, showing also increased thermal stabilities with low loadings of epoxy-grafted-GO [15]. Similarly, nylon 6-grafted-GO/nylon 6 composites prepared using an *in-situ* polymerization method were reported to improve the modulus of the compounds by 139% with the incorporation of 0.015 wt.% of the grafted GO in the polymer matrix, as well as to increase the degradation and the melting temperatures of the polymer [16].

Improving the compatibility and interface between carbon nanomaterials and thermoplastic polymers is challenging due to the polymers' low polarity and passive surface chemistry. However, an approach for preparing PMMA-functionalized-exfoliated graphene (GPMMA) via in situ free radical polymerization can be found in the literature [17]. In that work, the PMMA/GPMMA composites showed simultaneously improved Young's modulus, tensile stress, elongation at break and thermal stability by adding only 0.5 wt.% of GPMMA, which the authors attribute to a good dispersion of GPMMA and strong interfacial adhesion between GPMMA and the PMMA matrix. Chemically reduced GO sheets grafted with PMMA polymer by an emulsion polymerization have been also recently reported to give an efficient stress-transfer in the composites through an improved dispersion and strong sheet/matrix interfacial interaction [18]. In a different work, PMMA chains were 'grafted from' the GO surface via atom transfer radical polymerization (ATRP), improving the toughness and thermal stability of the composites through strong interfacial interactions between filler and host polymer [19]. These reported 'grafting from' methods (*i.e. in-situ* polymerization and ATRP) do not offer an easy up-scalability though, with alternative 'grafting to' methods emerging as more appropriate for industrial applications. Only one report [20] on the chemical grafting of PMMA chains to the surface of GO, leading to increases in Young's modulus and yield strength of a polyvinylidene fluoride and an acrylonitrile butadiene styrene polymer blend through an improved compatibility between the immiscible pair of polymers and improved stress-transfer at their interface, has been reported so far to the best of our knowledge. Crucially, almost all the studies discussed above report the modification of GO derived materials, which have intrinsic functional groups on their surfaces that facilitates the grafting reactions. It is important for industrial applications that alternative grafting techniques are developed for commercially available GNPs, since GNPs are significantly cheaper than GO and have no need for reduction to remove defects and restore the electrical and thermal conductivities.

We report here the first, highly scalable 'grafting to' method for the fabrication of covalently-bonded, polymer-functionalized commercially available graphene. This grafting is achieved by combining a diazonium addition reaction to give –$NH_2$ terminated GNPs (–$NH_2$-GNPs) with an amidation between the –$NH_2$ terminated GNPs and PMMA chains, to give PMMA grafted GNPs (PMMA-NH-GNPs). This process is specifically designed to reinforce PMMA composites through enhanced interfaces between filler and polymer in which the 'grafted' polymer chains are acting as 'bridges', connecting strongly both components of the composite and creating a 'continuity' between filler and matrix. GNPs/PMMA and PMMA-NH-GNPs/PMMA composite systems at loadings from 0.5 wt.% to 5 wt.% were prepared using the solvent method. The microstructure, thermal, mechanical and electrical properties of these composites were evaluated and related to the different surface chemistry of the GNPs and, thus, to the different graphene/polymer interfaces.

## 2. Experimental.

*2.1 Synthesis of $NH_2$ terminated graphene nanoplatelets ($NH_2$-GNPs)*

$NH_2$ terminated graphene platelets ($NH_2$-GNPs) were synthesized by a nucleophilic substitution reaction through a spontaneous diazonium coupling reaction. In a typical reaction, 150 mg of graphene nanoplatelets (GNP-M25 from XG Sciences, with lateral dimension and thickness of ~25 µm and ~6 nm, respectively, quoted by the manufacturer) was initially suspended in 125 ml of acetonitrile ($CH_3CN$) together with 1.5 g of p-phenylene diamine and heated up to 60 °C under mechanical stirring. Once this temperature was reached, 2 mL of isoamylnitrite were added to the mixture and left to react under mechanical stirring for 24h. The resultant solid was then separated by vacuum filtration, rinsed first three times with $CH_3CN$ and then three times with EtOH and left to dry at room temperature. This material was labelled as $NH_2$-GNPs. (Both reaction time and reactants ratio were previously optimized, details in Fig.S1, SI).

*2.2 Synthesis of PMMA grafted graphene nanoplatelets (PMMA-NH-GNPs)*

$NH_2$-GNPs and PMMA powders (50/85 mg) were dispersed together in chloroform ($CHCl_3$) and the mixture was stirred at 70 °C for 24 h. After that time, the reaction mixture was cooled down to room temperature, filtered to collect the resultant powder, washed with 50 mL of $CHCl_3$ and dried. This material was labelled as PMMA-NH-GNPs.

*2.3 (PMMA-NH-GNPs/PMMA) composite preparation*

The appropriate amount of filler (PMMA-NH-GNPs or GNPs) and 1 g of PMMA were dissolved in 15 mL of $CHCl_3$ using mechanical stirring at room temperature for 30 minutes to prepare composite materials (PMMA-NH-GNPs/PMMA or GNPs/PMMA) at various loadings from 0.5 to 5 wt.%. Composite films were prepared by depositing these dispersions on a Teflon dish using the solvent-casting method, followed by complete removal of the solvent and peeling-off of the composite films. Specimens with the desired shapes and sizes were cut out of these composite films for characterization.

*2.4 Characterization of the graphene materials and the composites*

Scanning Electron Microscopy (SEM) was used to characterize the microstructure of the composite materials using a Zeiss Ultra 55 FEG-SEM. Differential scanning calorimetry (DSC) was employed for the determination of the glass transition temperature ($T_g$) of the neat PMMA and the composites using a TA Q2000 DSC. Samples with a mass of approximately 10 mg were heated to 200 °C at a rate of 10 °C /min and held there for 3 minutes, in order to erase any thermal history. Subsequently, the samples were cooled to 25 °C at a rate of 10 °C /min. The samples were then heated at 10 °C/min in order to clearly observe the $T_g$.

X-ray photoelectron spectroscopy (XPS) was carried out using an Axis Ultra Hybrid spectrometer (Kratos Analytical, Manchester, United Kingdom), using monochromated Al Kα X-ray radiation at 1486 eV (10 mA emission at 15 kV, 150 W), under ultra-high vacuum at a base pressure of 1 x $10^{-8}$ mbar. Samples were pressed onto conductive copper tape. A charge neutraliser was used to remove any differential charging at the sample surface. Calibration of the binding energy (BE) scale was performed using the C 1s photoelectron peak at ~ 284.5 eV for graphitic carbon. Spectral deconvolution was performed using CASAXPS (www.casaxps.com) with Shirley-type backgrounds. Graphitic carbon was fit with an asymmetric peak shape analogous to those used to fit metallic peaks, due to the high conductivity of this carbon species. Graphitic carbon also exhibits a signature broad spectral feature between approximately 290-295 eV (a BE region where no other carbon species are expected), associated with excitation of the π- π* transition, which was fitted with one broad peak. While the choice of asymmetry factors for graphitic carbon is not well prescribed, they were optimized to fit the C1s spectrum for unmodified GNPs (where the spectra showed a minimal amount of other carbon species), and this was then applied to all spectra and found to yield satisfactory fits for all the C1s spectra consistently. This self-consistency between the spectra allowed for the relative amounts of graphitic carbon compared to other carbon species between unmodified and modified GNPs to be obtained. Other carbon species required for adequate fitting of the spectra included hydrocarbon at ~ 284.8 eV (fixed at an offset in BE of 0.3 eV from graphitic carbon, given the close proximity of the peak positions), C-N at approximately 285.5 eV, C-O at approx. 286.5 eV, C=O at ~ 289 eV, and finally the broad shakeup feature above 290 eV.

Thermogravimetric analysis (TGA) was performed at a heating rate of 10 °C/min under nitrogen using a TGA Q500 (TA Instruments). Stress-strain curves were obtained from dog-bone shaped specimens using an Instron 1122 machine, using a tensile rate of 0.05 mm/min and a load cell of 500 N. Micromechanical characterization was performed to follow interfacial stress-transfer in the nanocomposites. The shift of the Raman bands of graphene can be followed during deformation and it has been used to evaluate stress-transfer from the polymer matrix to the graphene reinforcement [4, 21]. *In-situ* Raman spectroscopy was performed on the nanocomposites deformed in four-point bending mode using a low-power 633 nm HeNe laser in the Renishaw 2000 spectrometer, with the Raman laser beam polarized parallel to the tensile axis. A strain gauge attached to the surface of the specimens was used to determine the strain applied. Atomic force microscope (AFM) images were acquired by using Dimension 3100 (Bruker) for the as-received nanoplatelets and NanoWizard 4 (JPK Instruments) for the functionalised graphene.

The impedances of the polymer and composites were tested on 6 mm x 6 mm x 0.2 mm films specimens using a PSM 1735 Frequency Response Analyzer from Newtons4th Ltd connected with Impedance Analysis Interface (IAI) at the range of frequencies from 1 to $10^6$ Hz. The specific conductivities ($\sigma$) of the materials were calculated from the measured impedances using:

$$\sigma(\omega) = |Y^*(\omega)| \frac{t}{A} = \frac{1}{Z^*} \times \frac{t}{A} \qquad \text{Equation (1)}$$

where $Y^*(\omega)$ is the complex admittance, $Z^*$ is the complex impedance, $t$ and $A$ are the thickness and cross section area of the sample, respectively.

## 3. Results and discussion

*3.1 Polymer chains grafting to the GNPs*

$NH_2$-terminated graphene nanoplatelets ($NH_2$-GNPs) prepared by a simple diazonium coupling were used here as 'platform' to graft the PMMA chains covalently to the nanoplatelets through an amidation between –$NH_2$ groups and the PMMA chains to render PMMA-NH-GNPs (PMMA undergoes this hydrolysis reaction due to the presence of methacrylic acid comonomer, as shown in the supplier's FT-IR), as schematically represented in Figure 1.

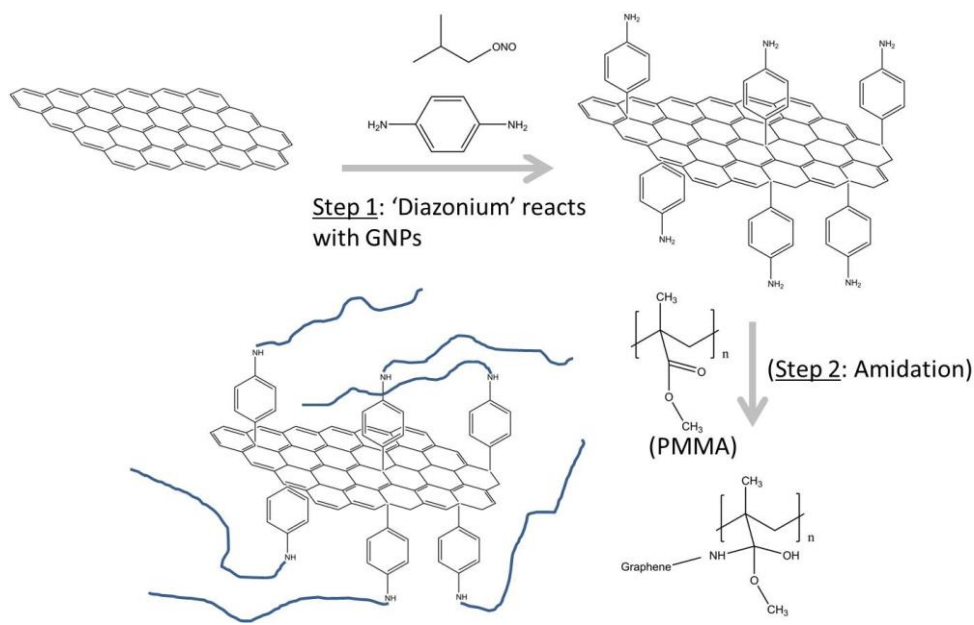

**Figure 1.** Scheme of the reaction developed to fabricate polymer functionalized GNPs.

The first step of the process involved a spontaneous diazonium coupling reaction performed on the as-received GNPs. It was previously reported by Dyke *et al.* [22] the ability of a diazonium salt to receive an electron from a carbon surface, liberating $N_2$ and forming the corresponding active radical. Since the formed active radical and the carbon 'substrate' are in close proximity, they eventually couple to each other and a covalent bond is formed between them. Raman spectroscopy, XPS and TGA analysis gave evidence of a successful diazonium coupling between the GNPs and the p-phenylene diamine, leading to the synthesis of $NH_2$-terminated graphene nanoplatelets through a similar mechanism to the one described by Dyke *et al.* [22].

The increased $I_D/I_G$ values revealed by Raman spectroscopy for the $NH_2$-GNPs relative to the starting GNPs (Fig. 2a) suggested a successful attachment of the –$NH_2$ groups to the nanoplatelets, since the functional groups are detected as defects on the carbon network by Raman spectroscopy. (The materials had been thoroughly rinsed, which further suggests covalent functionalization rather than just surface absorption). TGA analysis performed on the $NH_2$-GNPs (Fig. 3a) showed a weight loss of ~2.7 wt.% between 160 °C and 210 °C, which we relate to the removal of the –$NH_2$ functional groups attached to the GNPs (*i.e.* the thermal decomposition of p-phenylene diamine), and further evidenced the successful attachment of the diamine groups to the GNPs. In order to probe that this attachment occurs through the formation of covalent bonds, XPS analysis is required. High resolution XPS C1s spectra (Fig. 2b-d) showed the presence of -C-N- for the $NH_2$-GNPs, whereas no N was found in the starting GNPs. (For further evidence, high resolution N1s XPS spectra are shown in Fig. S2, SI). The amount of N incorporated on the GNPs during the diazonium coupling was found ~2.34 at.%, which corresponds to a N/C ratio of ~1/40 and suggests a degree of functionalization of 1 aniline group per ~40 carbons (survey XPS spectra and atomic composition shown in the SI).

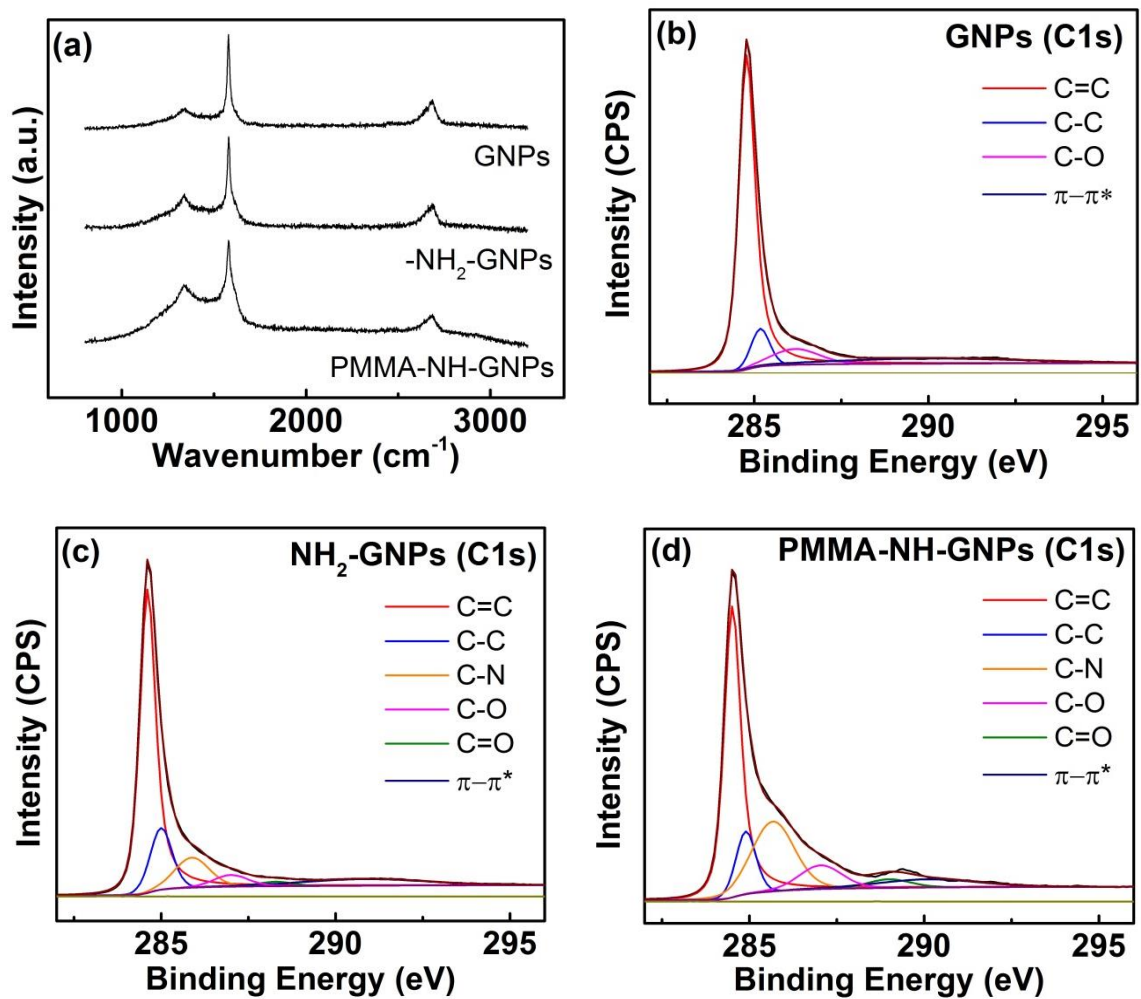

**Figure 2.** Raman spectra of the GNPs, -NH$_2$-GNPs and PMMA-NH-GNPs (a). High resolution C1s XPS spectra of the GNPs (b), NH$_2$-GNPs (c) and PMMA-NH-GNPs (d).

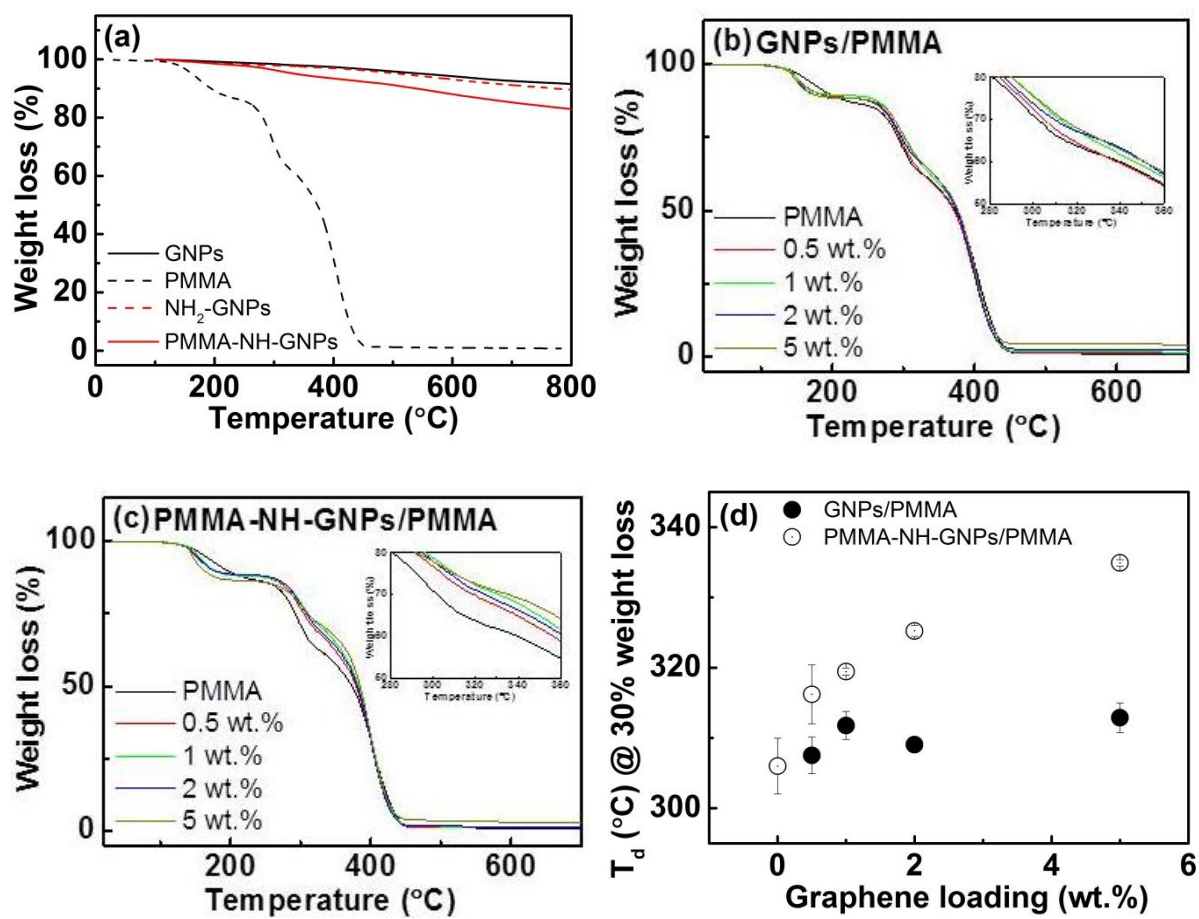

**Figure 3.** Thermogravimetric curves for the graphene materials (a) and composite systems (b, c) at different loadings (zooms inserted). Variation of the $T_d$ with loading for the composite systems (d).

On the second step of the chemical procedure PMMA chains were 'grafted to' the GNPs through an amidation between the attached –$NH_2$ groups and the PMMA chains to give PMMA-NH-GNPs (Fig. 1). Evidence of a successful grafting of the PMMA chains to the GNPs was given by Raman spectroscopy, XPS and TGA (Raman and TGA can only give evidence of a successful functionalization of the GNPs, whereas XPS gives evidence of a covalent attachment). Raman spectra showed an increase in $I_D/I_G$ relative to that found for the as-received GNPs (Fig. 2a), which suggested successful grafting of the polymer chains, since they are detected as defects on the carbon network by Raman spectroscopy (the slightly higher $I_D/I_G$ found for the PMMA grafted GNPs relative to the $NH_2$-GNPs must be related to the incorporation of defects and/or some additional functional groups or polymer chains physically adsorbed onto the surface of the flakes). In agreement with the Raman results, TGA analysis revealed a degree of polymer grafting of ~3.8 wt.% (given by the difference in weight loss found for the PMMA-NH-GNPs relative to GNPs at ~400 °C, which is the PMMA degradation temperature, Fig. 3a). XPS results revealed increased amounts of C-N (due to the covalent bonding between PMMA chains and $NH_2$-GNPs), as well as increased amounts of C-O, C=O and –COO (from the PMMA chains) for the PMMA-NH-GNPs (at.% determined by XPS compiled in Table S1; high resolution XPS N1s spectra shown in Fig.S2, SI), which gave strong evidence of an amidation between the -$NH_2$ functional groups covalently bonded to the GNPs and the PMMA chains as the 'grafting to' mechanism. (As a control experiment, an identical reaction was performed on the as-received GNPs with no -$NH_2$ groups attached and no PMMA grafting could be observed, Table S1 in the SI, which further evidences that the suggested 'grafting to' mechanism happens through the -$NH_2$ groups). In addition, XPS analysis revealed the incorporation of ~2 atoms of O per atom of N in the $NH_2$-GNPs during the grafting reaction, which corresponds to the attachment of one polymer chain per -$NH_2$, *i.e.* a degree of grafting of one polymer chain per ~40 atoms of C. In the field of composites it is important to gain control on the degree of the functionalization/grafting of the surface of the filler in order to 'design' the optimal graphene-polymer interface (a combination of a strong affinity between graphene and polymer and weak flake-flake interactions is required). We expect that the grafting of PMMA chains on the surface of the GNPs will enhance the compatibility/affinity between filler and polymer through chemical interactions due to a similar nature between the grafted polymer chains and the polymer matrix. The idea of controlling the level of grafting by controlling the level of oxidation of the starting graphene nanoplatelets should be kept in mind for future work.

*3.2 GNPs/PMMA and PMMA-NH-GNPs/PMMA composites*

Both as received unmodified GNPs and the PMMA grafted GNPs (*i.e.* PMMA-NH-GNPs) were incorporated into a PMMA matrix using a solvent casting method at loadings from 0.5 to 5 wt.%. The microstructure, thermal, mechanical and electrical properties of the composites were evaluated and related to both the dispersion of the flakes in the host polymer matrix and the strength of the graphene-polymer interface.

*3.2.1 Microstructure*

The microstructure of specimens of neat PMMA and of the two series of composites at different loadings were evaluated by SEM, and representative micrographs of the fracture sections are presented in Fig.4 (SEM micrographs of the top surface of the specimens are shown in SI, Fig. S3).

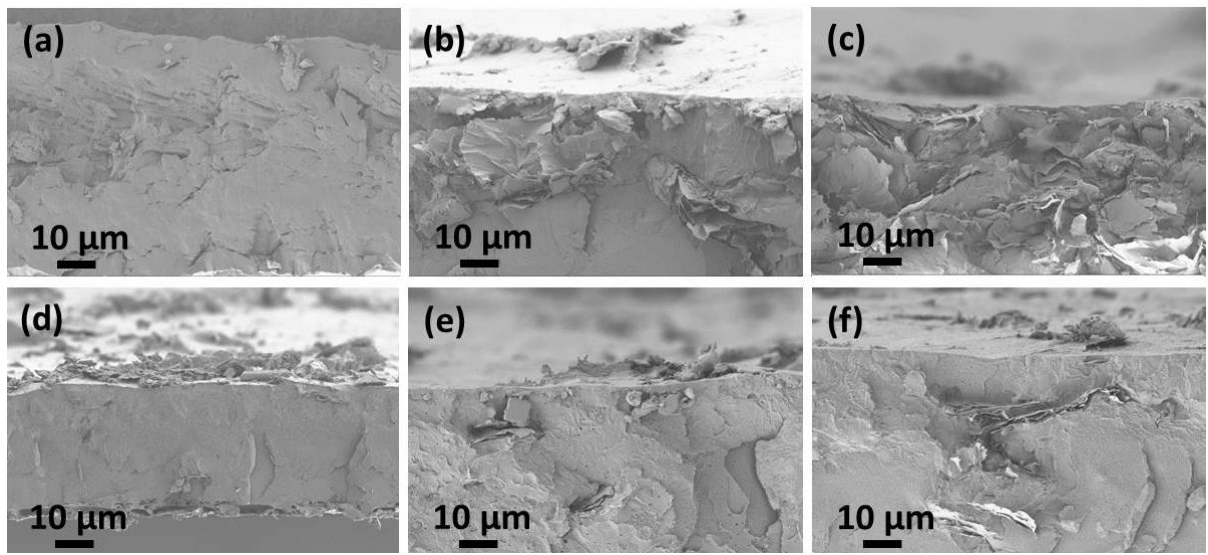

**Figure 4.** SEM micrographs of the fracture surface of: (a) PMMA, (b) GNPs/PMMA-0.5 wt.%, (c) GNPs/PMMA-5wt.%, (d) PMMA-NH-GNPs/PMMA-0.5 wt.%, (e) PMMA-NH-GNPs/PMMA-2 wt.%, (f) PMMA-NH-GNPs/PMMA-5 wt.%.

Fig.4 reveals a smooth surface for the neat PMMA specimen, whereas rougher surfaces were found for the composites at the studied loadings, with this 'roughness' increasing with loading of both modified and unmodified nanofillers. Fig.4(a-c) shows the presence of agglomerates of GNPs for the GNPs/PMMA composites at all loadings studied, from a loading as low as 0.5 wt.% and showing increasing amounts of agglomerates at higher loadings. (In thin composite films, below 1mm thick, prepared using the solvent-casting method, the effect of a different density between filler and polymer must be also contributing to this non-homogeneous distribution observed by SEM). However, for the PMMA-NH-GNPs/PMMA composites (Fig.4(d-f)) much lower levels of agglomeration were observed at all the loadings studied relative to the GNPs/PMMA ones with identical loadings. (At 0.5 wt.% content of PMMA-NH-GNPs, Fig.4d, practically no agglomerates could be found, whereas only a few agglomerates were observed at 5 wt.% loading, Fig.4f). SEM micrographs revealed, thus, improved dispersions for the PMMA grafted GNPs based composite system relative to the unmodified GNPs based one at all loadings studied. (This finding is supported by optical microscopy images of composites films at low loadings, as shown in Fig. S7). We relate this observation to an enhanced interaction/compatibility between graphene and host polymer after grafting PMMA chains to the GNPs. (We consider that this enhanced compatibility plays an important role in overcoming the different density issue between filler and polymer commented above, and it is expected to have important implications in the mechanical performance of the overall composites).

In addition, high-resolution SEM micrographs (Fig.S4) show the few-layered structure typical of the GNPs employed here in one flake embedded into the host polymer, which suggested no exfoliation of the nanoplatelets taking place during the process of functionalization, in agreement with X-ray diffraction performed on both as-received and chemically-modified GNPs (Fig.S5).

*3.2.2 Thermal properties and non-oxidative thermal stability of the composites*

DSC analysis was performed on the two series of composites (GNPs/PMMA and PMMA-NH-GNPs/PMMA) and the results are shown in Fig.5. The values found for the glass transition temperature ($T_g$) of all the composites prepared were higher than that found for the pure polymer, and they were found to increase with loading up to 5 wt.%. In general, increased values of $T_g$ for the nanocomposites with respect to neat polymer are typically related to the presence of attractive interactions or enhanced interfacial adhesion between the polar functional groups of the filler and the polymer, due to a chemical bond or other type of interaction between the host polymer and the surface of the filler [23-25].

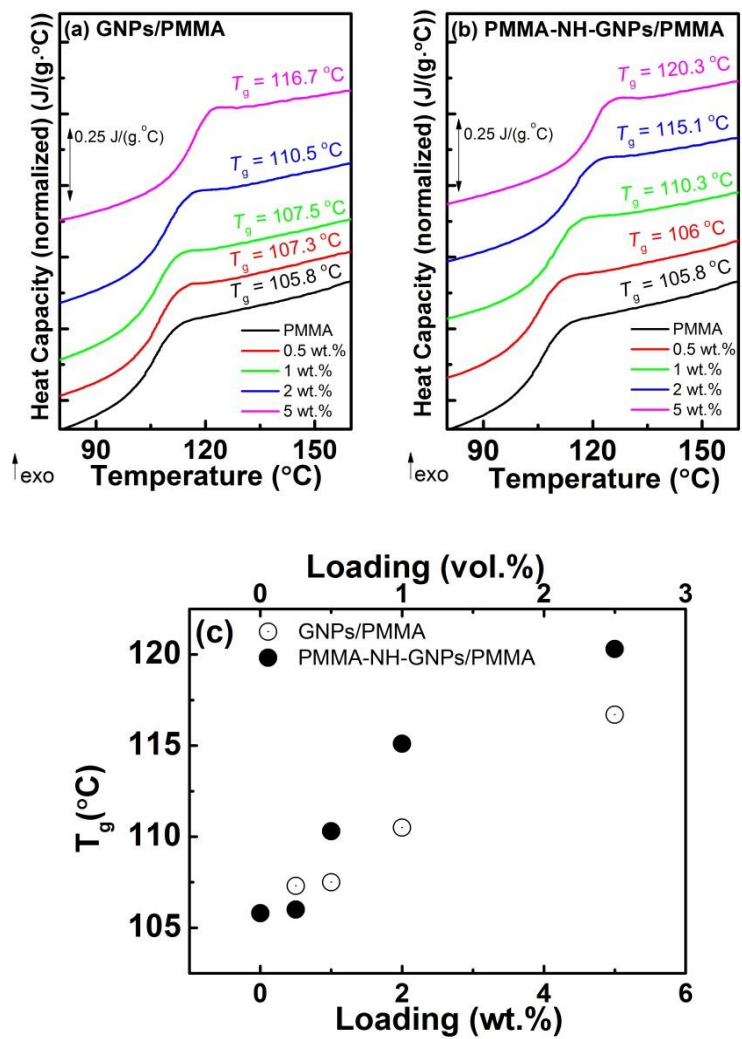

**Figure 5**. DSC of the GNPs/PMMA (a) and the PMMA-NH-GNPs/PMMA (b) series of composites. Variation of the $T_g$ with loading of filler (c).

At 0.5 wt.% loading the values for $T_g$ found for both composite systems were relatively close to each other, whereas at higher loadings the PMMA-NH-GNPs/PMMA system showed considerably higher $T_g$ values, with the maximum difference between both systems observed at loadings of 2 wt.%. (The $T_g$ reached maximum increases of 10 °C and 15 °C for 5 wt.% loadings of unmodified GNPs and PMMA-NH-GNPs, respectively, Fig.5c). We attribute the higher $T_g$ found for the polymer grafted GNPs based system to better dispersions and enhanced interfacial interactions between graphene and host polymer relative to the unmodified GNPs, which becomes more evident with increasing filler contents. Similar increases (~15 °C) in the $T_g$ of polystyrene (PS) were found for PS-grafted-GO/PS composites prepared by ATRP [26], whereas a maximum increment of ~24 °C was reported for polymer-grafted-GO/polyimide composite films prepared by *in-situ* polymerization [27]. This behaviour is attributed to the uniformly dispersed large aspect ratio graphene sheets with strong interfacial adhesion between filler and polymer, in total agreement with our observations.

Thermogravimetric analysis of both composite systems was carried under a nitrogen atmosphere in order to investigate how the chemical modification of the surface of the GNPs influences the non-oxidative thermal stability of the composites relative to the neat polymer, and the results are shown in Fig.3. Figs.3(b,c) show similar weight loss curve profiles for polymer and composites, with important enhancements of the thermal stability of the polymer observed by adding the studied nanofillers. (The three-step thermal degradation observed for the polymer and the composites corresponds to a mixture of poly(methyl methacrylate) and other methyl methacrylate copolymers). Furthermore, TGA analysis revealed higher values for the thermal degradation temperature of the PMMA ($T_d$, defined here as the temperature at 30% weight loss) with increasing loading of both fillers (Fig.3d). The improvement observed in the thermal stability of the polymer must be attributed to the formation of a high aspect ratio, heat-resistant graphene network in the host polymer matrix that acts as a barrier inhibiting the emission of small gaseous molecules and restraining the attack of free radicals generated during the thermal decomposition of the polymer, slowing thus the non-oxidative thermal degradation of the nanocomposites compared with the neat polymer [7, 28]. For the composites prepared with PMMA-NH-GNPs this effect is considerably more pronounced relative to the ones prepared with unmodified GNPs ($T_d$ of the polymer was increased by ~29 °C with the addition of 5 wt.% loading of PMMA-NH-GNPs, whereas it was observed to increase only by ~7 °C for an identical loading of unmodified GNPs), showing superior thermal stabilities. We attribute this observation to the existence of stronger interfaces between filler and host polymer for the modified GNPs, facilitating the capture of the radicals. These results are in agreement with other works reporting improved thermal stabilities of epoxy-grafted-GO/PC [15] and PMMA-grafted-GO/PMMA [19] composites, which the authors also attribute to stronger interfaces achieved through polymer grafting of the graphene nanofillers.

*3.2.3 Mechanical properties of the composites (Tensile testing).*

The mechanical properties for both series of composites were evaluated by tensile testing. The results show a glassy polymeric behaviour for the neat polymer and all the composites prepared here (Fig.6). The addition of as-received non-functionalized GNPs into the PMMA matrix led to a very poor reinforcement of the matrix at all studied loadings (only an increase in modulus by ~7% could be observed at 0.5 wt.%, with deterioration of the mechanical performance at higher contents). However, the incorporation of PMMA-grafted GNPs into the polymer was found to lead to important reinforcements, with a maximum increase in the modulus of the polymer of 24% found for the optimal 2 wt.% loading. No further reinforcement was observed at higher loadings though due to the formation of agglomerates of PMMA-NH-GNPs, as revealed by SEM, Fig.4. (The formation of agglomerates of graphene through π-π interactions and van der Waals forces has been widely observed and reported in similar polymer composites).

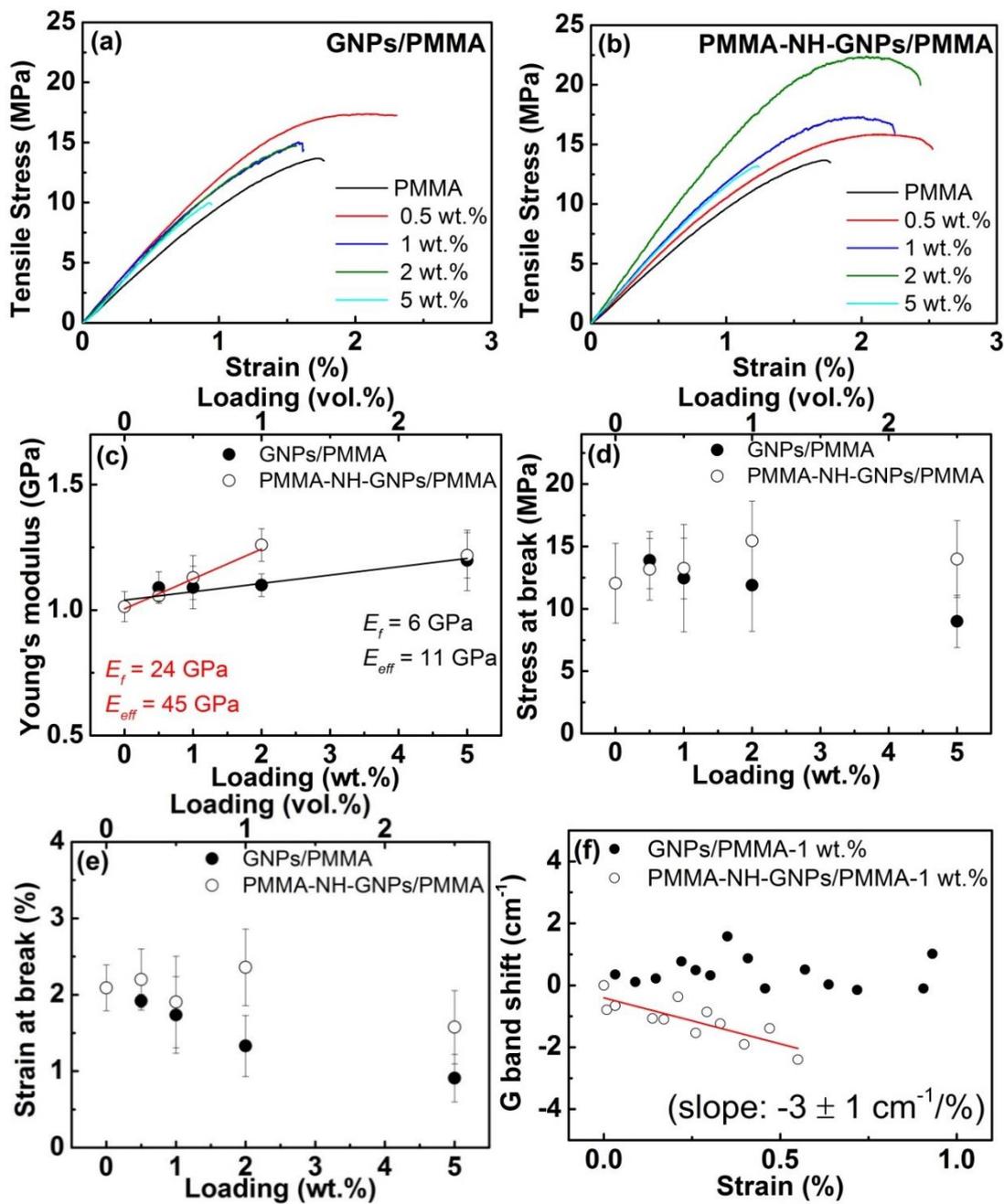

**Fig. 6.** Stress-strain curves obtained for GNPs/PMMA (a) and PMMA-NH-GNPs/PMMA (b) composite systems. Variation of the Young's modulus (c), stress at break (d) and strain at break (e) with loading for the two systems studied. Raman G band shift with strain applied to the 1 wt.% loading composites (f). (Errors were calculated from measuring at least seven specimens per loading for the tensile testing and two for the Raman bending).

Since a linear increase of modulus was found up to the optimal 2 wt.% loading for the PMMA grafted GNPs system, the rule of mixtures could be applied to determine the effective modulus ($E_f$) of the fillers in the nanocomposites using:

$$E_c = E_f V_f + E_m V_m \qquad \text{Equation (2)}$$

where $E_c$ is the Young's modulus of the composite, $E_f$, $E_m$ are the Young's modulus of the filler and matrix, respectively, and $V_f$ and $V_m$ are the volume fractions of the filler and the matrix, respectively, within the nanocomposite (with $V_f + V_m = 1$). The rule of mixtures has been widely employed to determine the effective modulus of graphene materials, $E_{\text{eff}}$, in similar systems [29], which could be calculated from $E_f$ using Equation (3):

$$E_f = E_{\text{eff}} \eta_0 \eta_l \qquad \text{Equation (3)}$$

where $\eta_0$ is the Krenchel orientation factor that depends on the average orientation of the particles [30] and $\eta_l$ is the length parameter which accounts for poor stress-transfer at the filler-matrix interface for particles with small lateral dimensions ($\eta_l = 1$ for perfect stress-transfer and $\eta_l = 0$ for no stress-transfer). For this calculation we employed $\eta_0 = 8/15$ [31] since SEM micrographs revealed a random orientation of the GNPs in the polymer matrix at all loadings studied of the two series of composites (results from polarized Raman spectroscopy, shown in the SI, revealed that very weak levels of orientation of the flakes in the matrix are promoted during the preparation of these composites, which can be thus ignored for these calculations), and $\eta_l = 1$ assuming a perfect stress-transfer. From the slopes of the lines up to the optimal loading in Fig.6c values of $E_f$ = 24 GPa and $E_{\text{eff}}$ = 45 ± 26 GPa were determined for the PMMA grafted nanoplatelets, which are higher than those found for the non-modified GNPs ($E_f$ = 6 GPa and $E_{\text{eff}}$ = 11 GPa).

Although the tensile testing results revealed good levels of reinforcement for the PMMA-NH-GNPs/PMMA composites (considerably superior than those found for the non-modified GNPs based composites), the effective modulus found here for this filler was still much lower than the 350 GPa theoretically predicted for the mechanical modulus of GNPs [32, 33]. Young *et al*. [32] recently developed a theory which explains why it is not possible to realise the promised 1 TPa or 350 GPa moduli for graphene or GNPs, respectively, in low modulus polymeric matrices. This theory predicts that the Young's modulus of a nanocomposite will be independent of the Young's modulus of the nanofiller, and highlights the importance of three structural parameters: (i) aspect ratio of the nanofiller, (ii) orientation of the nanofiller, and (iii) strength of the interface with the matrix (the theory shows that a strong nanofiller-matrix interface leads to good stress-transfer and hence better reinforcement than a weak interface). Even though the effective modulus found here for the polymer-grafted-GNPs is relatively low, the enhanced dispersions and strong filler-polymer interfaces achieved by grafting polymer chains on the reinforcements was found to lead to important levels of reinforcement, which suggests the importance of modifying chemically the interface between filler and host polymer to achieve very good mechanical performance, in agreement with the theory commented above.

Interestingly, a progressive increase of both strain (~13%) and stress at break (~28%) with loading was also observed up to the optimal 2 wt.% loading of the PMMA-grafted-graphene, whereas no improvement on either the stress or the strain at break was found when the non-functionalized GNPs were incorporated into the matrix at all studied loadings (Fig.6), similarly to previously reported graphene/polymer or GO/polymer composites [29], which were typically related to the formation of agglomerates at the interface acting as failure points. The enhanced interface between GNPs and PMMA achieved through polymer grafting of the flakes must be the responsible for the enhanced stress and strain at break observed here, in agreement with other work reporting significant enhancements in the mechanical properties of polymer matrices with the addition of small amounts of polymer-grafted-GO materials [14-16, 18-20, 34].

In general terms, the mechanical behaviour of both unmodified and PMMA-grafted GNPs were found similar to each other at low loadings (~0.5 wt.%), whereas at higher loadings the PMMA-grafted-GNPs system showed a superior mechanical performance, in agreement with our results from DSC (the values for $T_g$ were higher for the grafted-GNPs/composites above 0.5 wt.% content). This observation suggested a relatively good dispersion of the unmodified GNPs at loadings below 0.5 wt.% (evidenced by SEM, Fig.4), with agglomeration and subsequent deterioration of the thermal and mechanical performance at higher contents. The superior mechanical performance observed for the PMMA-grafted-GNPs at loadings above 0.5 wt.% must be attributed to a combination of better dispersions of the flakes in the host polymer relative to the unmodified GNPs (up to 2 wt.% loading) and stronger flake-polymer interfaces. The different surface chemistry of the flakes must be the responsible for this findings, since no other significant modifications on the structure of the flakes could be found (X-Ray diffraction revealed that no exfoliation happened during the chemical procedure described in this work, whereas AFM characterization showed no significant variation on the dimensions of the flakes during the process, as shown in SI). AFM showed the presence of thicker flakes after PMMA grafting though, which we relate to the formation of a PMMA 'layer' covering the surface of the flakes, which must be the responsible for the improved filler-polymer interface observed.

In general, the composites based on PMMA-NH-GNPs presented here show better mechanical performance relative to previously reported works which use unfunctionalized GNPs, graphene oxide, CNTs or other inorganic nanofillers. The use of unmodified GNPs in PMMA was probed to lead to important increases in the Young's modulus of the polymer (78 %) up to high loadings (~20 wt.%) [6], whereas the use of graphene with a chemically modified surface was probed to effectively reduce the amount of graphene filler required to achieve comparable levels of reinforcement. (Indeed, ~1 wt.% was typically found as the optimal loading for GO/PMMA composites [7], reaching ~37 % improvement in modulus and 22 % in strength (relative to pure polymer), with deterioration of the mechanical properties of the overall composite at higher loadings due to the formation of agglomerates of GO in the matrix). The presence of functionalities on the surface of graphene promotes simultaneously both strong graphene-polymer and graphene-graphene interactions: at loadings below 1 wt.% the graphene-polymer interactions dominate the system, leading to good dispersions of the GO in the polymer matrix and, thus, to enhanced mechanical performance, whereas at higher loadings the flake-flake interactions are dominating the system and the formation of agglomerates through π-π interactions deteriorate the mechanical properties of the composites. Grafting PMMA chains on the surface of the GNPs provides an improved compatibility between filler and polymer while reducing the flake-flake interactions, which leads to increased optimal loading (2 wt.%). Although the increment in modulus (24%) is slightly lower than that observed for GO (37%), the tensile stress and strain are considerably higher than those observed for the GO/PMMA system. (Composites based on graphene and other related materials typically show a very brittle behaviour and they do not typically increase the stress at break either, due to the formation of small agglomerates in the matrix acting as failure points [29]. However, the PMMA-grafted-GNPs reported here led to increments of both these properties).

The results reported here are also superior to those typically observed when using CNTs as fillers in thermoplastics, mainly due to the difficulty to achieve good dispersions of CNTs in the polymer matrix at high loadings combined with a lower processability, relative to graphene related materials. In general, CNTs-based PMMA composites require higher loadings of filler ($\geq$ 3 wt.%) to observe increased moduli and strengths, while elongation at break is not enhanced due to the brittle nature of these materials [35, 36]. Strength of the composite was found to improve at lower CNTs loading using functionalized nanotubes compared to unfunctionalized nanotubes [37, 38], due to improved dispersions, although no improvements in the elongation at break were found. Finally, regarding the use of nanoparticles with different morphology to reinforce PMMA, a lot of work has been done in developing chemical strategies to modify the compatibility between PMMA and $SiO_2$ nanoparticles, for example, in order to improve the dispersion and the mechanical performance through enhanced compatibility and polymer-filler interactions. These strategies have been reported to successfully lead to improved elastic modulus and scratch resistance [39], surface hardness, flexural strength, impact strength as well as tensile strength and modulus [40, 41] in thin films. However, this enhanced mechanical performance still has to be probed in bulk composite materials to be attractive from an industrial point of view. This discussion evidences that the particle size and dispersion definitely play an important role by tuning the surface load transfer, which affects the final materials load capacity. Graphene materials lead to considerably better mechanical performances (including modulus, strength and strain at break) due to higher surface area and 2D nature, relative to CNTs and other inorganic nanoparticles, through improved interfaces and larger surfaces for chemical modification.

*3.2.4 Stress-induced Raman Band Shifts.*

The nature of the interface between GNPs and the host polymer in the two composite systems studied was evaluated by Raman spectroscopy. In order to understand the observed mechanical behaviour and the mechanics of these reinforcements, we evaluated the stress-transfer at the graphene-polymer interface, using Raman spectroscopy. Fig.6f shows the position of the Raman G band determined in the nanocomposites as a function of strain for 1 wt.% loaded composite specimens. The rate of the shift of the Raman bands with the applied strain in a nanocomposite is known to be proportional to the effective Young's modulus of a graphene-based reinforcement [4, 21]. The knowledge of the Grüneisen parameter for graphene [42], which revealed a Raman G-band rate shift of -30 cm$^{-1}$/% strain for a sheet of graphene with a Young's modulus of 1050 GPa [43], allows us to determine the effective Young's modulus of our graphene filler from Raman spectroscopy ($E_R$), using Equation (4):

$$E_R = -\frac{d\omega(G)}{d\varepsilon} \times \frac{1050}{-30} \text{ GPa} \qquad \text{Equation (4)}$$

where $-\frac{d\omega(G)}{d\varepsilon}$ is the G band rate shift in cm$^{-1}$ per % applied strain.

Hence the slope of the lines observed in Fig.6f for the PMMA-NH-GNPs/PMMA composite sample is an indication of the effective modulus and stress-transfer from the polymer to the graphene occurring at the interface (no shift was found for the GNPs/PMMA sample at identical loading). The G band shift rate of 3 ±1 cm$^{-1}$/% observed for the PMMA-NH-GNPs/PMMA sample led to values of effective modulus of ~100 ± 40 GPa using Equation (4), which is in agreement with the results obtained experimentally using tensile testing in *Section 3.2.3*. (This is about twice the value of the mechanical modulus, which is reasonable as the Raman tends to pick out large aligned flakes and tend to generate higher modulus values [29]). There is no need to use the Krenchel orientation factor [30] in this case as the polarised laser beam detects predominately those flakes that are aligned parallel to the tensile axis [44]. This observation from polarized Raman gives a direct evidence of a stronger interface between graphene and polymer after polymer chain grafting relative to the unmodified GNPs based system, which is in agreement with the experimental evidence found above from the thermal and mechanical characterization performed on these composite systems. In addition, high resolution SEM images of the fracture surface of these two composite materials (Figure S4, SI) revealed that the flakes remain more embedded in the polymer matrix after fracture with less voids at the interface between the functionalized graphene flakes and the polymer matrix, relative to the as-received GNPs, which gives further evidence of a stronger interface/interaction between the functionalized GNPs and the polymer matrix. Due to the similar nature of the polymer chains grafted on the GNPs and the polymer matrix chemical interactions between the chemically modified filler and the matrix are expected, which must be responsible for this better compatibility and stronger interface.

*3.2.5 Electrical properties of the GNPs/PMMA and PMMA-NH-GNPs/PMMA composites.*
The electrical conductivity of both GNPs/PMMA and PMMA-NH-GNPs/PMMA series of composites at loadings from 0.5 to 10 wt.% were tested and the results are shown in Fig.S6, in the SI. The GNPs/PMMA system behaved as a percolated system, showing a percolation threshold ~3.5 wt.% and reaching maximum conductivities of ~0.04 S/cm. The PMMA-NH-GNPs/PMMA system, however, did not show electrical conductivity at any loading up to 5 wt.%. In order to explain this observation we should consider two important points: (i) the disruption of the $Csp^2$ network typical of graphenic/graphitic materials, and (ii) the dispersion of the nanofiller in the polymer matrix. The functionalization of graphene related materials, such as the grafting of PMMA chains onto the surface of the GNPs reported here, is known to generate a 'disruption' of the $Csp^2$ network. Some C=C bonds (which are responsible for the electrical conductivity typically measured for graphene/graphite) open in order to incorporate these functionalities, which generates a combination of $Csp^2$ and $Csp^3$. This generation of $Csp^3$ domains typical of functionalized graphene materials leads to lower intrinsic electrical conductivities relative to the non-functionalized ones (in which the $Csp^2$ domains are dominating). That is the main reason why it was not possible to measure electrical conductivity for the PMMA-NH-GNPs/PMMA composite system at any studied loading (up to 10 wt.% loading). In addition to this disruption of the $Csp^2$ network, the enhanced dispersion found for the modified nanoplatelets must be also playing an important role in the electrical behaviour. It is known that very good dispersions of flakes in the matrix tend to deteriorate the conductivity due to an increasing difficulty for the creation of a conductive path throughout the composite material. The good dispersion observed for the functionalized flakes must be contributing to the observed deterioration of the electrical conductivity. A combination of the disruption of their Csp$^2$ network and the enhanced dispersions of the flakes in the host matrix achieved with this chemical modification was probed to transform electrically conductive composite materials into insulating ones.

## 4. Conclusions

PMMA grafted graphene nanoplatelets were successfully prepared using a simple diazonium coupling reaction to give NH$_2$-terminated flakes followed by an amidation reaction between the NH$_2$ groups and PMMA chains (PMMA-NH-GNPs) in order to improve the compatibility of graphene with a PMMA host matrix through stronger filler-polymer interactions. Raman spectroscopy, XPS and TGA analysis revealed successful covalent attachment, corresponding to 1 polymer chain per ~40 atoms of C. The polymer grafted graphene was found to render important improvements on the mechanical performance of PMMA composites (including improvements on the elastic modulus, stress and strain at break) up to the optimal 2 wt.% loading, whereas poor reinforcement was observed with the addition of unmodified GNPs at all studied loadings. In addition, enhanced thermal stabilities and increased $T_g$ up to high loadings (5 wt.%) were found for the PMMA-NH-GNPs/PMMA system relative to both pure polymer and the GNPs/PMMA system. We attribute this higher mechanical and thermal performance of the polymer-grafted-GNPs system relative to the unmodified graphene based one to improved dispersions of the flakes in the host polymer matrix and stronger graphene/polymer interfaces, as revealed by SEM and Raman spectroscopy, respectively, caused by the polymer chains grafted to the graphene acting as 'bridges' and strongly connecting graphene and host polymer in the composites. This work offers the possibility of improving and controlling the physical properties and functionality of graphene/polymer composites through chemically tuning the graphene/polymer interface, which will broaden considerably the range of applications in the field of graphene based nanocomposites. The method developed here shows great potential for industrial applications due to an easy fabrication procedure, high productivity (> 70 wt.%) and the use of commercially available GNPs as starting material. Further work and investigation is still required, however, to optimize the process, maximizing the yield and minimizing both the use of solvents and the generation of waste products.

**Acknowledgements.** This project has received funding from the European Union's Horizon 2020 research and innovation programme under grant agreement No 785219. IAK also acknowledges the Royal Academy of Engineering and Morgan Advanced Materials. In addition, we would like to acknowledge *The Henry Royce Institute for Advanced Materials,* funded through EPSRC grants EP/R00661X/1 and EP/P025021/1, for access to the atomic force microscope.

**References**

[1] Park S, Ruoff RS. Chemical methods for the production of graphenes. Nat Nanotechnol. 2009;4(4):217-224.

[2] Geim AK. Graphene: Status and prospects. Science. 2009;324(5934):1530-1534.

[3] Geim AK, Novoselov KS. The rise of graphene. Nat Mater. 2007;6(3):183-191.

[4] Gong L, Kinloch IA, Young RJ, Riaz I, Jalil R, Novoselov KS. Interfacial stress transfer in a graphene monolayer nanocomposite. Adv Mater. 2010;22(24):2694-2697.

[5] Vallés C, Abdelkader AM, Young RJ, Kinloch IA. Few layer graphene-polypropylene nanocomposites: The role of flake diameter. Faraday Discussions. 2014;173:379-390.

[6] Vallés C, Abdelkader AM, Young RJ, Kinloch IA. The effect of flake diameter on the reinforcement of few-layer graphene-PMMA composites. Compos Sci Technol. 2015;111:17-22.

[7] Vallés C, Kinloch IA, Young RJ, Wilson NR, Rourke JP. Graphene oxide and base-washed graphene oxide as reinforcements in PMMA nanocomposites. Compos Sci Technol. 2013;88:158-164.

[8] Bahr JL, Yang JP, Kosynkin DV, Bronikowski MJ, Smalley RE, Tour JM. Functionalization of carbon nanotubes by electrochemical reduction of aryl diazonium salts: A bucky paper electrode. J Am Chem Soc. 2001;123(27):6536-6542.

[9] Kariuki JK, McDermott MT. Nucleation and growth of functionalized aryl films on graphite electrodes. Langmuir. 1999;15(19):6534-6540.

[10] Liu MC, Duan YX, Wang Y, Zhao Y. Diazonium functionalization of graphene nanosheets and impact response of aniline modified graphene/bismaleimide nanocomposites. Mater Design. 2014;53:466-474.

[11] Park OK, Hahm MG, Lee S, Joh HI, Na SI, Vajtai R, et al. In Situ Synthesis of Thermochemically Reduced Graphene Oxide Conducting Nanocomposites. Nano Lett. 2012;12(4):1789-1793.

[12] Yu DS, Kuila T, Kim NH, Lee JH. Enhanced properties of aryl diazonium salt-functionalized graphene/poly(vinyl alcohol) composites. Chem Eng J. 2014;245:311-322.

[13] Balazs AC, Emrick T, Russell TP. Nanoparticle polymer composites: Where two small worlds meet. Science. 2006;314(5802):1107-1110.

[14] Guan LZ, Wan YJ, Gong LX, Yan D, Tang LC, Wu LB, et al. Toward effective and tunable interphases in graphene oxide/epoxy composites by grafting different chain lengths of polyetheramine onto graphene oxide. J Mater Chem A. 2014;2(36):15058-15069.

[15] Shen B, Zhai WT, Tao MM, Lu DD, Zheng WG. Chemical functionalization of graphene oxide toward the tailoring of the interface in polymer composites. Compos Sci Technol. 2013;77:87-94.

[16] Nguyen L, Choi SM, Kim DH, Kong NK, Jung PJ, Park SY. Preparation and characterization of nylon 6 compounds using the nylon 6-grafted GO. Macromol Res. 2014;22(3):257-263.

[17] Wang JL, Shi ZX, Ge Y, Wang Y, Fan JC, Yin J. Solvent exfoliated graphene for reinforcement of PMMA composites prepared by in situ polymerization. Mater Chem Phys. 2012;136(1):43-50.

[18] Gong LX, Pei YB, Han QY, Zhao L, Wu LB, Jiang JX, et al. Polymer grafted reduced graphene oxide sheets for improving stress transfer in polymer composites. Compos Sci Technol. 2016;134:144-152.

[19] Gonalves G, Marques PAAP, Barros-Timmons A, Bdkin I, Singh MK, Emami N, et al. Graphene oxide modified with PMMA via ATRP as a reinforcement filler. J Mater Chem. 2010;20(44):9927-9934.

[20] Kar GP, Biswas S, Bose S. Tailoring the interface of an immiscible polymer blend by a mutually miscible homopolymer grafted onto graphene oxide: outstanding mechanical properties. Phys Chem Chem Phys. 2015;17(3):1811-1821.

[21] Frank O, Mohr M, Maultzsch J, Thomsen C, Riaz I, Jalil R, et al. Raman 2D-band splitting in graphene: Theory and experiment. ACS Nano. 2011;5(3):2231-2239.

[22] Dyke CA, Tour JM. Unbundled and highly functionalized carbon nanotubes from aqueous reactions. Nano Lett. 2003;3(9):1215-1218.

[23] Qiao R, Catherine Brinson L. Simulation of interphase percolation and gradients in polymer nanocomposites. Compos Sci Technol. 2009;69(3-4):491-499.

[24] Bansal A, Yang H, Li C, Cho K, Benicewicz BC, Kumar SK, et al. Quantitative equivalence between polymer nanocomposites and thin polymer films. Nat Mater. 2005;4(9):693-698.

[25] Ellison CJ, Torkelson JM. The distribution of glass-transition temperatures in nanoscopically confined glass formers. Nat Mater. 2003;2(10):695-700.

[26] Fang M, Wang KG, Lu HB, Yang YL, Nutt S. Covalent polymer functionalization of graphene nanosheets and mechanical properties of composites. J Mater Chem. 2009;19(38):7098-7105.

[27] Abbasi S, Carreau PJ, Derdouri A, Moan M. Rheological properties and percolation in suspensions of multiwalled carbon nanotubes in polycarbonate. Rheologica Acta. 2009;48(9):943-959.

[28] Papageorgiou DG, Terzopoulou Z, Fina A, Cuttica F, Papageorgiou GZ, Bikiaris DN, et al. Enhanced thermal and fire retardancy properties of polypropylene reinforced with a hybrid graphene/glass-fibre filler. Compos Sci Technol. 2018;156:95-102.

[29] Papageorgiou DG, Kinloch IA, Young RJ. Mechanical properties of graphene and graphene-based nanocomposites. Prog Mater Sci. 2017;90:75-127.

[30] Krenchel H. Fibre Reinforcement. Akademisk Forlag: Copenhagen, 1964.

[31] Li Z, Young RJ, Wilson NR, Kinloch IA, Vallés C, Li Z. Effect of the orientation of graphene-based nanoplatelets upon the Young's modulus of nanocomposites. Compos Sci Technol. 2016;123:125-133.

[32] Young RJ, Liu MF, Kinloch IA, Li SH, Zhao X, Vallés C, et al. The mechanics of reinforcement of polymers by graphene nanoplatelets. Compos Sci Technol. 2018;154:110-116.

[33] Gong L, Young RJ, Kinloch IA, Riaz I, Jalil R, Novoselov KS. Optimizing the reinforcement of polymer-based nanocomposites by graphene. ACS Nano. 2012;6(3):2086-2095.

[34] Liao WH, Yang SY, Wang JY, Tien HW, Hsiao ST, Wang YS, et al. Effect of Molecular Chain Length on the Mechanical and Thermal Properties of Amine-Functionalized Graphene

Oxide/Polyimide Composite Films Prepared by In Situ Polymerization. ACS Appl Mater Inter. 2013;5(3):869-877.

[35] Pal K. Effect of different nanofillers on mechanical and dynamic behavior of PMMA based nanocomposites. Compos Commun. 2016;1:25-28.

[36] Ben David O, Banks-Sills L, Aboudi J, Fourman V, Eliasi R, Simhi T, et al. Evaluation of the Mechanical Properties of PMMA Reinforced with Carbon Nanotubes - Experiments and Modeling. Exp Mech. 2014;54(2):175-186.

[37] Mathur RB, Pande S, Singh BP, Dhami TL. Electrical and mechanical properties of multi-walled carbon nanotubes reinforced PMMA and PS composites. Polym Composite. 2008;29(7):717-727.

[38] Madeshwaran SR, Kwon JK, Cho JW. Functionalized multi-walled carbon nanotubes with hyperbranched aromatic polyamide for poly(methyl methacrylate) composites. Fiber Polym. 2013;14(2):182-187.

[39] Qu M, Meth JS, Blackman GS, Cohen GM, Sharp KG, Van Vliet KJ. Tailoring and probing particle-polymer interactions in PMMA/silica nanocomposites. Soft Matter. 2011;7(18):8401-8408.

[40] Hong RY, Fu HP, Zhang YJ, Liu L, Wang J, Li HZ, et al. Surface-modified silica nanoparticles for reinforcement of PMMA. J Appl Polym Sci. 2007;105(4):2176-2184.

[41] Stojanovic D, Orlovic A, Markovic S, Radmilovic V, Uskokovic PS, Aleksic R. Nanosilica/PMMA composites obtained by the modification of silica nanoparticles in a supercritical carbon dioxide-ethanol mixture. J Mater Sci. 2009;44(23):6223-6232.

[42] Mohiuddin TMG, Lombardo A, Nair RR, Bonetti A, Savini G, Jalil R, et al. Uniaxial strain in graphene by Raman spectroscopy: G peak splitting, Grüneisen parameters, and sample orientation. Physical Review B. 2009;79(20):205433.

[43] Li Z, Young RJ, Kinloch IA. Interfacial stress transfer in graphene oxide nanocomposites. ACS Appl Mater & Inter. 2013;5(2):456-463.

[44] Young RJ, Kinloch IA, Gong L, Novoselov KS. The mechanics of graphene nanocomposites: A review. Compos Sci Technol. 2012;72(12):1459-1476.